\documentclass[11pt]{article}

\usepackage[a4paper,margin=1in]{geometry}
\usepackage[T1]{fontenc}
\usepackage[utf8]{inputenc}   % si compilas con pdflatex; si usas lualatex/xelatex, puedes quitarlo
\usepackage{lmodern}
\usepackage{microtype}
\usepackage{cite}

\usepackage{amsmath,amssymb,bm}
\usepackage{graphicx}
\usepackage{placeins}
\usepackage{xcolor}
\usepackage{hyperref}
\hypersetup{
  colorlinks=true,
  linkcolor=blue,
  citecolor=blue,
  urlcolor=blue
}
\usepackage{color}

\usepackage{fancyhdr}
\newcommand{\PaperTitle}{Optical pulse-induced quantum geometric waves in graphene}
\newcommand{\AuthorLine}{Luis F.\ C\'ardenas-Castillo$^{1}$ and Wei Chen$^{2}$}
\newcommand{\AffilLine}{$^{1,2}$\textit{Department of Physics, PUC-Rio, 22451-900 Rio de Janeiro, Brazil}}
\newcommand{\EmailLine}{\texttt{$^{1}$lcardenasc@aluno.puc-rio.br}}
\newcommand{\DateLine}{(\today)}

\makeatletter
\newcommand{\InlineTOC}{%
  \@starttoc{toc}%
}
\makeatother

\newcommand{\MakeFrontPage}{%
  \thispagestyle{empty} % hide page number on first page (but it still counts as page 1)

  \begin{center}
    {\Large\bfseries \PaperTitle\par}
    \vspace{0.8em}
    {\normalsize \AuthorLine\par}
    \vspace{0.35em}
    {\small \AffilLine\par}
    \vspace{0.25em}
    {\small \EmailLine\par}
    \vspace{0.6em}
    {\small \DateLine\par}
  \end{center}

  \vspace{0.8em}
  \hrule
  \vspace{0.9em}

  {\bfseries \large{Abstract}}\par
  \begingroup
  \small

  We show that, under a short optical pulse, the quantum metric of Bloch states in the momentum-time $(k_x,k_y,t)$ of graphene becomes dynamic and exhibits a wave-like behavior near Dirac points. This quantum metric wave reflects the Floquet-band structure caused by the pulse, as revealed by solving the time-dependent Schr\"{o}dinger equation assuming that correlations and out-of-equilibrium effects can be ignored. The momentum and temporal components of the metric have very distinct time dependence that persists even after the pulse has passed. In addition, the pulse also generates a Berry curvature wave that is otherwise absent in static graphene. The time-dependent electron densities in conduction and valence bands also give arise to a Fisher information wave that constitutes part of the quantum metric wave, and is readily measurable by pump-probe experiments.
  
  \par
  \vspace{0.4em}
  \textbf{Keywords:} quantum metric; Berry curvature; graphene; Fisher information.
  \endgroup

  \vspace{0.9em}
  \hrule
  \vspace{0.8em}

  {\bfseries \large{Contents}}\par
  \vspace{-0.3em}
  \begingroup
  \small
  \InlineTOC
  \endgroup

  \vspace{0.7em}
  \hrule

}

\begin{document}

% Front page (single column)
\MakeFrontPage

% ---------------------------
\section{Introduction}

The quantum geometry of Bloch states in the momentum space of solids has emerged as an important topic that is currently under intensive investigation. This notion of quantum geometry arises when one considers the overlap between neighboring Bloch states in a $D$-dimensional momentum space, whose leading-order expansion defines a quantum geometric tensor. The real part of the tensor gives the quantum metric\cite{Provost80}, while the imaginary part gives the Berry curvature that encodes geometric phases and many topological and transport properties~\cite{Berry84,Xiao10}. Particularly in semiconductors and insulators, this quantum metric coincides with the optical transition matrix elements of interband transitions\cite{Ozawa18,Ahn22}, and thus a wide variety of dielectric and optical properties, such as optical conductivity, dielectric function, refractive index, absorption coefficient, reflectance, etc, are all determined by the quantum metric\cite{Komissarov24,Chen25_optical_marker}, which has stimulated a great deal of interest on the quantum metric. 

%Furthermore, in topological materials, a metric-curvature correspondence dictates that the Berry curvature that integrates to the Chern number in 2D and the Berry connection that integrates to the winding number in 1D are both directly proportional to the square root of the determinant of the quantum metric\cite{Ma13,Ma14,Yang15,Piechon16,Panahiyan20_fidelity,vonGersdorff21_metric_curvature}. As a result, several optical properties are manifestations of topological order, such as the roughly frequency-independent absorbance of graphene as a topological charge\cite{deSousa23_graphene_opacity}. 

Beyond in static systems, quantum geometry in time-dependent systems has also been an intriguing issue that manifests many unique features that are absent in their static counterparts\cite{Grandi10,Grandi10_2,Pang17,Jafari20,Wu24,Diaz25}. Firstly, because the time dependence of the quantum state is no longer described by a simple dynamic phase factor, time itself serves as an additional parameter that promotes the geometry to a $(D+1)$-dimensional momentum-time quantum geometry. The quantum metric in this case is in a way analogous to the spacetime metric in classical gravity, although one should keep in mind that momentum-time is a Euclidean manifold, unlike the Lorentzian manifold of spacetime. Secondly, the time dependence of the quantum state also renders the quantum metric time-dependent. In most situations, the dynamic quantum metric oscillates in a wavy fashion, of which we call a quantum-metric wave, serving as an analog of the gravitational wave in general relativity, although many components of the quantum metric wave are highly nonlinear and do not behave like a simple sinusoidal wave.

To elaborate these unique features in time-dependent systems, and also motivated by the optical effects related to static quantum metric, we resort to the quantum geometry of graphene in the setup of pump-probe experiments\cite{Dawlaty08,Wang10,Breusing11,Strait11,Winnerl11,Tielrooij13,Johannsen13,Gierz13,Brida13,Winnerl13,Plotzing14,Wagner14,Gierz15}. In this type of experiment, a short laser pulse of sub-picosecond duration is applied to graphene, causing nonequilibrium excitations of electrons from the valence to conduction bands, which then relax back to equilibrium via rather complicated many-body interactions. This subject has been investigated intensively owing to the rich out-of-equilibrium physics involved, such as the Floquet band formation, light-induced pseudospin textures, and pump-probe photoemission~\cite{Sentef15,Schuler21,Merboldt25}. To capture the essential features of the dynamic quantum geometry by the simplest approach, we solve the time-dependent Schr\"{o}dinger equation in momentum space under the influence of the pulse\cite{Grifoni98}, while ignoring all the complications coming from out-of-equilibrium and many-body effects. We anticipate that such a simplified approach can capture the evolution of quantum metric during a relatively weak pulse, while it is unable to describe the relaxation back to equilibrium long after the pulse. Using this simple approach, we quantify the momentum and temporal dependence of the quantum metric in the three-dimensional (3D) momentum-time $(k_{x},k_{y},t)$, revealing intriguing wave-like patterns that coincide with the Floquet bands. Remarkably, we find that the laser pulse also generates a Berry curvature wave in the momentum-time mainly distributed around the Dirac point, which is absent in an otherwise static graphene, pointing to the significance of optical engineering of quantum geometric properties.

To relate the quantum metric wave to experimental observables, we further invoke the aspects of information geometry into the problem\cite{Amari16}. We elaborate that from the time-varying valence and conduction electron densities, a Fisher information matrix\cite{Fisher25} can be introduced to quantify the information that can be extracted from the electron density as a time-varying probability mass function. The Fisher information matrix constitutes part of the dynamic quantum metric\cite{Facchi10}, and can be extracted from the time-varying valence and conduction electron densities measured by time- and angle-resolved photoemission spectroscopy (tr-ARPES), suggesting that the information geometry part of the dynamic quantum geometry can be readily measured.

\section{Driven two-band model for graphene under optical pulse}
\label{sec:model}

\subsection{Driven honeycomb-lattice Hamiltonian}

In this section, we formulate the coherently driven two-band description of graphene and specify the pure-state interband dynamics underlying the quantum geometry. We start from a monolayer graphene described within a nearest-neighbor tight-binding model on the honeycomb lattice\cite{CastroNeto09}. In the sublattice basis $(A,B)$, the static Bloch Hamiltonian is
\begin{equation}
    H_0(\boldsymbol{k}) = 
    \begin{pmatrix}
        0&d_1(\boldsymbol{k})-id_2(\boldsymbol{k})\\
        d_1(\boldsymbol{k})+id_2(\boldsymbol{k})&0
    \end{pmatrix}
    \label{static_graphene_Hk}
\end{equation}
with
\begin{equation}
    \begin{split}
        d_1(\boldsymbol{k})&=\gamma[\cos\boldsymbol{k}\cdot\boldsymbol{a}_{1} + \cos\boldsymbol{k}\cdot\boldsymbol{a}_{2} + 1]  \\
        d_2(\boldsymbol{k})&=\gamma[\sin\boldsymbol{k}\cdot\boldsymbol{a}_{1} + \sin\boldsymbol{k}\cdot\boldsymbol{a}_{2}].
    \end{split}
    \label{d_vector_graphene}
\end{equation}
Here $\gamma=2.6$ eV is the nearest-neighbor hopping amplitude\cite{Kochan17}, $\boldsymbol{a}_{1}=a\left(3/2,\sqrt{3}/2\right)$ and $\boldsymbol{a}_{2}=a\left(3/2,-\sqrt{3}/2\right)$ are the primitive lattice vectors, and $a=1.42$ $ \mathring{A}$ is the carbon-carbon distance. The conduction and valence band energies are $E_c(\boldsymbol{k})=+d(\boldsymbol{k})$ and $E_v(\boldsymbol{k})=-d(\boldsymbol{k})$, respectively, where $d(\boldsymbol{k}) = \sqrt{d^2_1(\boldsymbol{k})+d^2_2(\boldsymbol{k})}$. 

%This form makes explicit the effective two-level structure of graphene near the Dirac points, with the pseudospin texture encoded in $d_1(\boldsymbol{k})$ and $d_2(\boldsymbol{k})$.

%At the exact Dirac point, \(d(\mathbf{k})=0\), the static conduction-valence basis is not uniquely defined. This is an isolated degeneracy of the static problem. In the numerical implementation, a small energy regularizer is used only to avoid division by zero when constructing the local static spinor frame, and the resulting time-dependent spinor is normalized after construction. This prescription fixes a local convention at the isolated degeneracy and does not affect the finite-\(\mathbf{k}\) resonant structures discussed below.

%The regularizer is not added to the Hamiltonian and does not open a physical gap; it only fixes the numerical representation of the static spinor frame at the isolated degeneracy.

%We therefore do not assign physical significance to the single grid point exactly at the static degeneracy; all conclusions concern the finite-\(\mathbf{k}\) resonant structures surrounding the valley.

The optical pump is incorporated through the Peierls substitution by means of a time-dependent vector potential $\boldsymbol{A}(t)=(A_x(t), A_y(t))=\hat{\boldsymbol{\epsilon}}V(t)$, where $\hat{\boldsymbol{\epsilon}}$ specifies the polarization direction, and $V(t)$ describes the sinusoidal oscillation and Gaussian envelope of the pulse shape 
\begin{equation}
    V(t) = \dfrac{A_0}{\sqrt{2\pi} \tau}\exp\left[- \dfrac{(t-t_0)^2}{2\tau^2}\right]\sin(\Omega t).
    \label{Vt_pulse_shape}
\end{equation}
The factor \(A_0\) fixes the overall strength of the vector potential, chosen such that the corresponding electric
field \(\mathbf{E}(t)=-\partial_t\boldsymbol{A}(t)\) has a total fluence\cite{Gierz13}
\begin{equation}
    F=\frac{\epsilon_0 c}{2}\int dt\, |\mathbf{E}(t)|^2=4.6~{\rm mJ\,cm^{-2}}.
\end{equation}
In Eq.~(\ref{Vt_pulse_shape}), $t_0$ specifies the pulse center, $\tau$ controls the envelope width, and $\Omega$ is the carrier frequency. Throughout the paper, we choose $\hbar\omega_{\text{pump}}=\hbar\Omega = 950~\mathrm{meV}$, $\tau=18 $ fs, $t_0=70$ fs and chemical potential $\mu=0$, to simulate a typical pump-probe experiment\cite{Gierz13}. 

%In addition, we will examine the dependence on the polarization by investigating three configurations of the vector potential $(A_{x}(t),A_{y}(t))$, namely polarization along the armchair direction $(V(t),0)$, zigzag direction $(0,V(t))$, and diagonal direction $(V(t)/\sqrt{2},V(t)/\sqrt{2})$.

The drive enters the tight-binding Hamiltonian through the Peierls substitution
\begin{equation}
    \theta_i(t) = \dfrac{e}{\hbar}\int^{\boldsymbol{r}+\boldsymbol{\delta}_i}_{\boldsymbol{r}}\boldsymbol{A}(t)\cdot d\boldsymbol{x}
\end{equation}

where $\boldsymbol{\delta_i}$ are the nearest-neighbor vectors of graphene:
\begin{equation}
    \boldsymbol{\delta}_1 = a\left(\dfrac{1}{2},\dfrac{\sqrt{3}}{2} \right),\hspace{0.2cm}\boldsymbol{\delta}_2 = a\left(\dfrac{1}{2},-\dfrac{\sqrt{3}}{2} \right),\hspace{0.2cm}\boldsymbol{\delta}_3 = a\left(-1,0 \right)
\end{equation}

Evaluating the line integral along the three bonds, we obtain
\begin{equation}
   \begin{split}   
    \theta_1(t)&=\alpha_P\left( \dfrac{1}{2}A_x(t) + \dfrac{\sqrt{3}}{2}A_y(t)\right) \\
    \theta_2(t)&=\alpha_P\left( \dfrac{1}{2}A_x(t) - \dfrac{\sqrt{3}}{2}A_y(t)\right) \\
    \theta_3(t) &=-\alpha_P A_x(t)
   \end{split}
\end{equation}
where $\alpha_P = ea/\hbar$, which modifies the ${\bf d}$-vector in Eq.~(\ref{d_vector_graphene}) by
\begin{equation}
    \begin{split}
        \tilde{d}_1(\boldsymbol{k},t) &= \gamma[\cos(\boldsymbol{k}\cdot\boldsymbol{a}_{1} - \theta_1(t)) + \cos(\boldsymbol{k}\cdot\boldsymbol{a}_{2} - \theta_2(t))+\cos(\theta_3(t))], \\
        \tilde{d}_2(\boldsymbol{k},t) &= \gamma[\sin(\boldsymbol{k}\cdot\boldsymbol{a}_{1} - \theta_1(t)) + \sin(\boldsymbol{k}\cdot\boldsymbol{a}_{2} - \theta_2(t))-\sin(\theta_3(t))].
    \end{split}
\end{equation}
Since many features revealed by our numerical results in the following sections are found to be robust against the direction of polarization, we will only present the results for the polarization along the armchair direction ${\boldsymbol A}=(A_{x}(t),0)$ for simplicity.

%\begin{figure}[h]
%    \centering
%    \includegraphics[width=8cm]{BZ_honeycomb_lattice.jpeg}
%    \caption{Brillouin zone of graphene, used in this manuscript. Here, $Q$ point is defined as the place in BZ where the transition energy $2d(\boldsymbol{k})$ equals the pump energy established as 950 meV and corresponds to a interband pumping resonance. This is obtained approximately in $q=0.049$. }
%    \label{fig:BZ_figure}
%\end{figure}

\subsection{Pulse-induced driven quantum tunneling in graphene}

In terms of the dimensionless $n_1(\boldsymbol{k}) = d_1(\boldsymbol{k})/d(\boldsymbol{k})$ and $n_2(\boldsymbol{k}) = d_2(\boldsymbol{k})/d(\boldsymbol{k})$ that satisfy $n^2_1(\boldsymbol{k}) + n^2_2(\boldsymbol{k}) = 1$, the conduction $|u_c(\boldsymbol{k})\rangle$ and valence $|u_v(\boldsymbol{k})\rangle$ band eigenstates of the static Hamiltonian $H_0(\boldsymbol{k})$ in Eq.~(\ref{static_graphene_Hk}) are
\begin{equation}
    |u_c(\boldsymbol{k})\rangle = \dfrac{1}{\sqrt{2}}
    \begin{pmatrix}
        1\\
        n_1(\boldsymbol{k}) + in_2(\boldsymbol{k})
    \end{pmatrix},\hspace{0.3cm}
    |u_v(\boldsymbol{k})\rangle = \dfrac{1}{\sqrt{2}}
    \begin{pmatrix}
        1\\
        -n_1(\boldsymbol{k}) - in_2(\boldsymbol{k})
    \end{pmatrix}.
\end{equation}
To analyze the driven dynamics, we follow the formalism of driven quantum tunneling to expand the time-dependent state by the static eigenstates\cite{Grifoni98}
\begin{equation}
    |\Psi(\boldsymbol{k},t)\rangle = c_1(\boldsymbol{k},t)e^{-id(\boldsymbol{k})t/\hbar}|u_c(\boldsymbol{k})\rangle + c_2(\boldsymbol{k},t)e^{+id(\boldsymbol{k})t/\hbar}|u_v(\boldsymbol{k})\rangle,
    \label{psikt_uc_uv}
\end{equation}
which evolves according to the time-dependent Schr\"odinger equation
\begin{equation}
    i\hbar\partial_t|\Psi(\boldsymbol{k},t)\rangle = H(\boldsymbol{k},t)|\Psi(\boldsymbol{k},t)\rangle.
\end{equation}
In terms of the two real functions that encode the pulse and momentum dependence
\begin{equation}
\begin{split}
    \Xi(\boldsymbol{k},t)&= \dfrac{1}{\hbar}[d(\boldsymbol{k})-\tilde{d}_1(\boldsymbol{k},t)n_1(\boldsymbol{k})-\tilde{d}_2(\boldsymbol{k},t)n_2(\boldsymbol{k})]  \\
    \Lambda(\boldsymbol{k},t)&=\dfrac{1}{\hbar}[\tilde{d}_1(\boldsymbol{k},t)n_2(\boldsymbol{k})-\tilde{d}_2(\boldsymbol{k},t)n_1(\boldsymbol{k})]
\end{split}
\end{equation}
the Schr\"odinger equation leads to the equations of motion for the amplitudes
\begin{equation}
\begin{split}
    \dot{c}_1(\boldsymbol{k},t) &=  i\Xi(\boldsymbol{k},t)c_1(\boldsymbol{k},t)-\Lambda(\boldsymbol{k},t)e^{+2id(\boldsymbol{k})t/\hbar}c_2(\boldsymbol{k},t),\\
    \dot{c}_2(\boldsymbol{k},t) &=-i\Xi(\boldsymbol{k},t)c_2(\boldsymbol{k},t)+\Lambda(\boldsymbol{k},t)e^{-2id(\boldsymbol{k})t/\hbar}c_1(\boldsymbol{k},t),
\end{split}
\label{c1c2_differential_eq}
\end{equation}
indicating that the function $\Xi(\boldsymbol{k},t)$ describes a drive-induced intraband phase modulation, whereas $\Lambda(\boldsymbol{k},t)$ controls the interband coupling. To draw relevance to pump-probe experiments, we assume that before the pulse kicks in at $t=0$, all the electrons are in the valence band states $|\Psi(\boldsymbol{k},t=0)\rangle = |u_v(\boldsymbol{k})\rangle$, leading to the initial condition for the complex amplitudes
\begin{eqnarray}
    ({\rm Re}\,c_1(\boldsymbol{k},0),{\rm Im}\,c_1(\boldsymbol{k},0),{\rm Re}\,c_2(\boldsymbol{k},0),{\rm Im}\,c_2(\boldsymbol{k},0))=(0,0,1,0).
\end{eqnarray}
Equation~(\ref{c1c2_differential_eq}) can then be solved numerically at each
momentum $\boldsymbol{k}$ on an \(N_k\times N_k\) grid, where we used $N_k=31$.

\section{Quantum geometry and information geometry in momentum-time}
\label{sec:tqgt}

%Within the driven two-band framework introduced in Sec.~2, the central geometric object is the quantum geometric tensor associated with the evolving spinor $|\Psi(\boldsymbol{k},t)\rangle$. Since the state is defined in the extended parameter space $(\boldsymbol{k},t)$, its geometry must be characterized not only in momentum space but also in the temporal and mixed time-momentum sectors. This section introduces that tensor and then focuses on the local valley response, where the pump-induced geometric structures are most pronounced. Global momentum-space maps of the metric and Berry curvature are provided in the Supplemental Material.

\subsection{Quantum geometric tensor in momentum-time}

Our goal is to examine the quantum geometry of the time-dependent state $|\Psi(\boldsymbol{k},t)\rangle$ in the momentum-time ${\boldsymbol\lambda}=(k_x,k_y,t)$ treated as a 3D Euclidean manifold. Since the state is properly normalized $|c_1(\boldsymbol{k},t)|^2+|c_2(\boldsymbol{k},t)|^2 = 1$ at any point in momentum-time, one can introduce the quantum geometric tensor of this state by
\begin{equation}
    Q_{\mu\nu}(\lambda)=
    \langle\partial_{\mu} \Psi(\lambda)|\partial_{\nu}\Psi(\lambda) \rangle
    - \langle \partial_{\mu}\Psi(\lambda)|\Psi(\lambda) \rangle
      \langle \Psi(\lambda) | \partial_{\nu}\Psi(\lambda)\rangle,
\label{tQGT_complete}
\end{equation}
where $\partial_{\mu}=\partial/\partial\lambda^{\mu}$ is the derivative on either the momentum or time calculated numerically by means of the finite difference on the grid. The real part of the tensor defines the time-dependent quantum metric
\begin{equation}
    g_{\mu\nu}(\lambda)=\mathrm{Re}\,Q_{\mu\nu}(\lambda),
\end{equation}
which characterizes the overlap of the time-dependent states on neighboring points on the momentum-time manifold (repeating indices are summed)
\begin{equation}
    |\langle\Psi(\boldsymbol{k},t)|\Psi(\boldsymbol{k}+\delta \boldsymbol{k},t+\delta t)\rangle|^2=
    1-g_{\mu\nu}(\boldsymbol{k},t)\delta\lambda^{\mu}\delta\lambda^{\nu} .
    \label{fidelity_expansion_metric}
\end{equation}
On the other hand, the imaginary part of the tensor gives a time-dependent Berry curvature,
\begin{equation}
    \Omega_{\mu\nu}(\lambda)=-2\,\mathrm{Im}\,Q_{\mu\nu}(\lambda),
\end{equation}
that is known to be absent in static graphene.

%of the normalized spinor on the \((k_x,k_y,t)\) grid. Before taking differences, neighboring spinors are locally phase-aligned as
%\[
%|\Psi'\rangle\rightarrow
%e^{-i\arg\langle\Psi|\Psi'\rangle}|\Psi'\rangle ,
%\]
%which removes arbitrary local \(U(1)\) phases from the finite-difference derivatives. For the local valley maps, the momentum derivatives are computed on the \(31\times31\) grid centered at \(K\), while the temporal derivative is evaluated from the spinors at \(t\) and \(t+\Delta t\), with \(\Delta t=0.05\,\mathrm{fs}\). The tensor is then computed from Eq.~(22).

\begin{figure}[h]
    \centering
    \includegraphics[width=0.8\columnwidth]{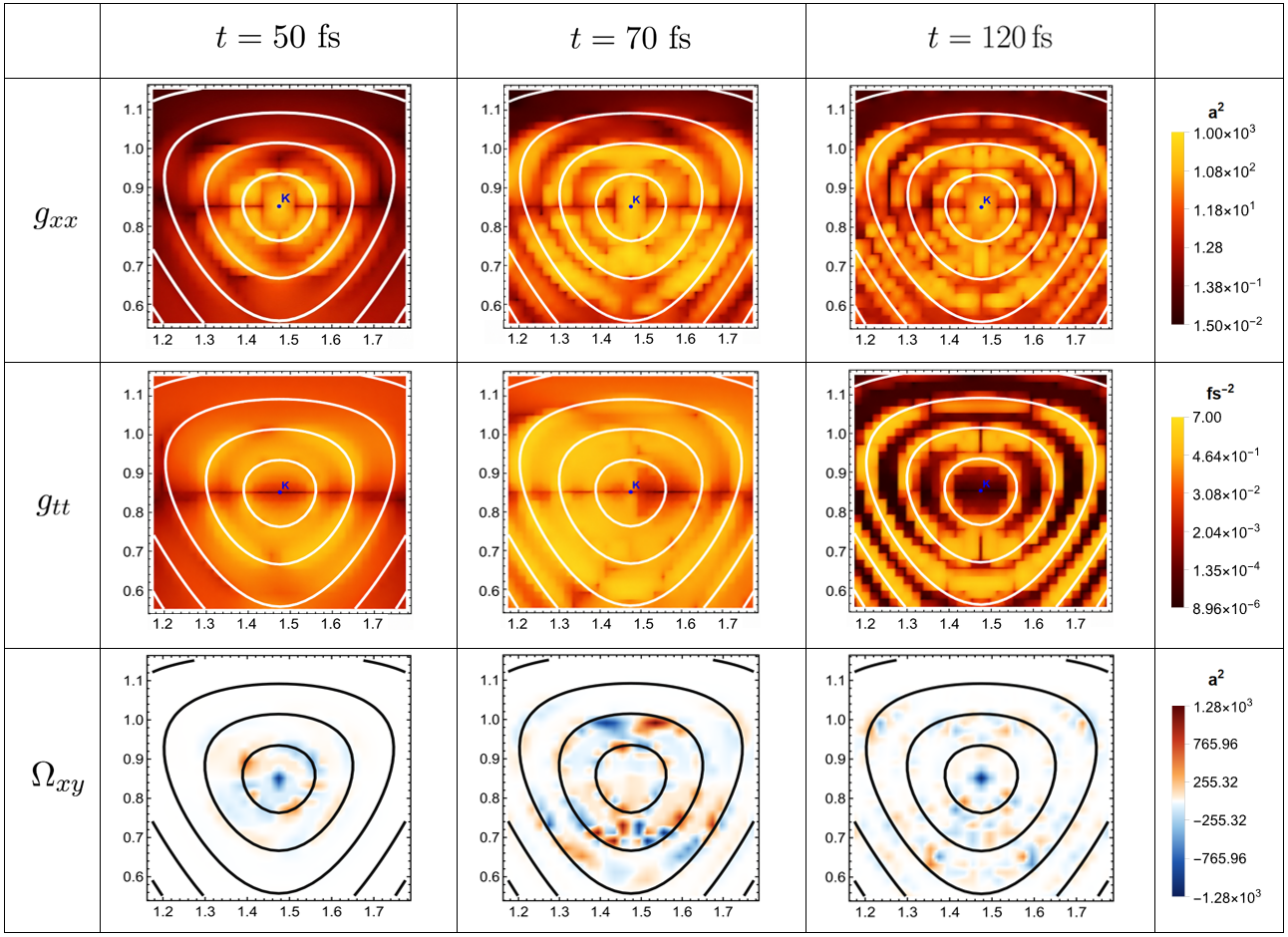}
    \caption{Momentum profile of the elements of the dynamic quantum geometric tensor near the \(K\) valley under the influence of an \(x\)-polarized optical pulse at the beginning of the pulse $t=50$fs, at the peak of the pulse $t=70$fs, and after the pulse $t=120$fs. The panels show the momentum-space quantum metric \(g_{xx}/a^2\), the temporal metric \(g_{tt}/{\rm fs}^{-2}\), and the Berry curvature \(\Omega_{xy}/a^2\), all of which manifest ring-like features that respect the Floquet bands labeled by white and black contours.}
    \label{fig:metric_waves_near_K}
\end{figure}

\subsection{Quantum geometric waves \label{sec:quantum_geometric_waves}}

Figure~\ref{fig:metric_waves_near_K} shows numerical results near the \(K\)-valley, where we present the momentum $g_{xx}$ and temporal $g_{tt}$ components of the quantum metric, as well as the momentum component of the Berry curvature \(\Omega_{xy}\). We present the results at the beginning of the pulse $t=50$ fs, close to the peak of the pulse $t=70$ fs, and right after the pulse $t=120$ fs. The momentum components \(g_{k_\mu k_\nu}\) and \(\Omega_{k_\mu k_\nu}\) are plotted in units of \(a^2\), while the temporal component \(g_{tt}\) is plotted in units of \({\rm fs}^{-2}\), and the data are presented in logarithmic scale in order to highlight the pattern in momentum space. 

All the geometric quantities in Fig.~\ref{fig:metric_waves_near_K} clearly display ring-like features near the $K$-valley, which we refer to as quantum geometric waves. A similar pattern also appears near the $K'$-valley, but we omit for simplicity. To further understand the origin of these rings, we plot the contour in momentum space where the optical excitation energy, i.e., energy difference between conduction and valence bands, matches multiples of the photon energy
\begin{equation}
E_{c}(\boldsymbol{k})-E_{v}(\boldsymbol{k})=2d(\boldsymbol{k})={\rm integer}\times\hbar\omega_{\rm pump},
\label{resonance_condition}
\end{equation}
presented as the white contours in Fig.~\ref{fig:metric_waves_near_K}. The coincidence of these resonance contours and the ring-like features of $\left\{g_{xx},g_{tt},\Omega_{xy}\right\}$ is evident, indicating that the rings are a finite-pulse analogue of Floquet sideband formation\cite{Sentef15}. In other words, the quantum geometry of time-dependent systems senses the formation of Floquet bands. Note that while the quantum metric $g_{xx}$ and $g_{tt}$ maintain roughly the same sign along the contour, the Berry curvature $\Omega_{xy}$ changes sign along the contour and has a much more complicated pattern around the $K$-valley. The momentum integration of the Berry curvature remains zero at all times, indicating no Chern number is induced by the pulse.

\begin{figure}[ht]
    \centering
    \includegraphics[width=0.8\columnwidth]{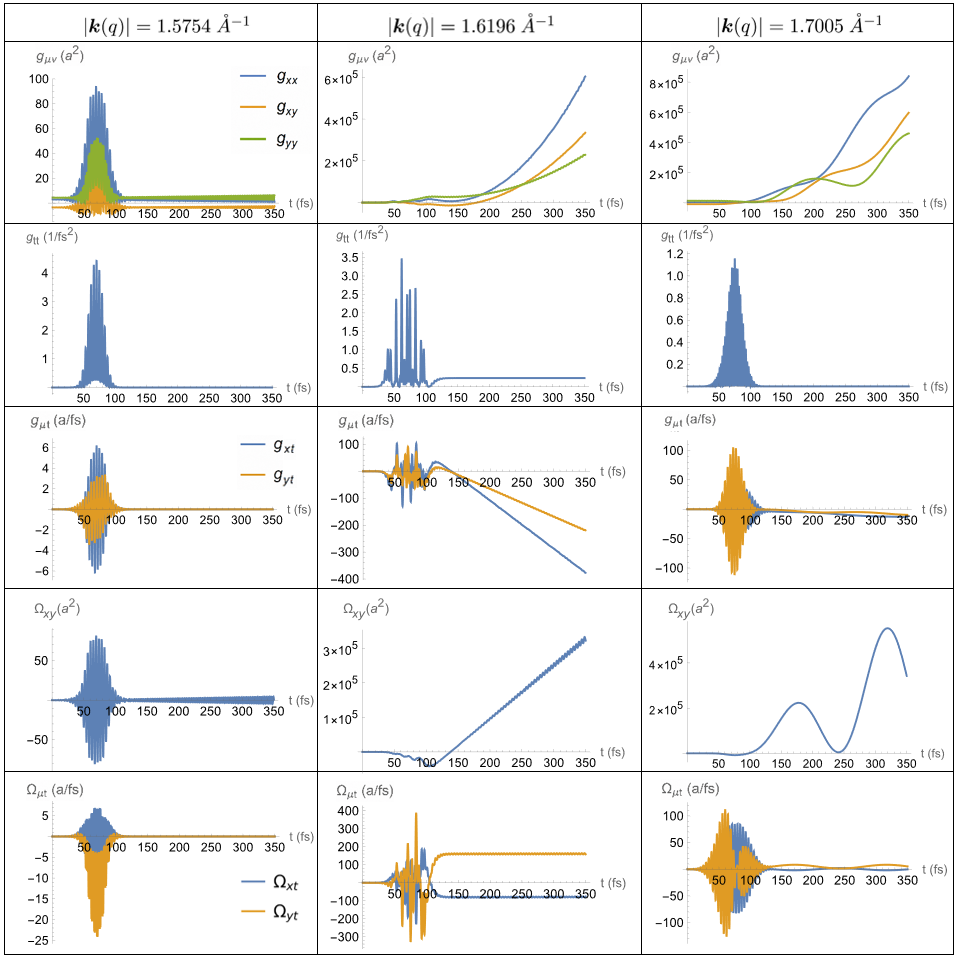}
    \caption{Time dependence of the dynamic quantum geometric tensor along the \(\Gamma-K\) direction caused by an \(x\)-polarized pulse, evaluated at below $|\boldsymbol{k}(q)|=1.5754\AA^{-1}$, exactly at $|\boldsymbol{k}(q)|=1.6196\AA^{-1}$, and above $|\boldsymbol{k}(q)|=1.7005\AA^{-1}$ the first resonant
condition $E_{c}(\boldsymbol{k})-E_{v}(\boldsymbol{k})=\hbar\omega_{\rm pump}$. We present the momentum component of the quantum metric $\left\{g_{xx},g_{xy},g_{yy}\right\}$ (first row), temporal component of the metric $g_{tt}$ (second row), mixed component of the metric $\left\{g_{xt},g_{yt}\right\}$ (third row), momentum component of the Berry curvature \(\Omega_{xy}\) (fourth row), and mixed component of the curvature $\left\{\Omega_{xt},\Omega_{yt}\right\}$ (fifth row). }
    \label{fig:tqgt_gammaK}
\end{figure}

%The same finite-difference and local phase-alignment procedure is used for the global reciprocal-space maps. For the boundary points of the displayed grid we use one-sided finite differences, while the internal consistency checks reported in the Supplemental Material are evaluated on the interior finite-difference grid.

%\subsection{Momentum-cut dynamics and the resonant point \texorpdfstring{$Q$}{Q}}

%We use two momentum-space samplings. The valley-resolved figures, used to analyze the ring-like structures near the Dirac point, are computed on a local grid centered at \(K\), covering \(|k_x-K_x|\le 0.3\,\mathrm{\AA}^{-1}\) and \(|k_y-K_y|\le 0.3\,\mathrm{\AA}^{-1}\). The polarization-resolved maps in Fig.~5 and Figs.~S1--S2 are computed on a uniform global reciprocal-space grid containing the first-Brillouin-zone region shown in the figures. In both cases the same grid size \(N_k=31\) is used.

%To resolve the temporal origin of the local structures, we examine the quantum geometric tensor along the $\Gamma-K$ direction. The characteristic point $Q$ is defined by the local resonance condition
%\begin{equation}
%    2d(\boldsymbol{k}_Q)=\hbar\omega_{\mathrm{pump}},
%\end{equation}
%so that the static interband splitting matches the pump photon energy. This point provides a natural reference for separating off-resonant and near-resonant dynamics.

To further understand the temporal dependence of the quantum geometric waves, we fix the momentum and investigate the evolution of these geometric quantities as a function of time. Specifically, we focus on the high symmetry line connecting the $\Gamma$ and $K$ points, and around the momentum that satisfies the first resonance condition $E_{c}(\boldsymbol{k})-E_{v}(\boldsymbol{k})=\hbar\omega_{\rm pump}$ according to Eq.~(\ref{resonance_condition}). In Fig.~\ref{fig:tqgt_gammaK}, we present the results of time dependence at momenta slightly larger than the first resonance and farther away from the $K$-point $|\boldsymbol{k}(q)|=1.5754\AA^{-1}$, exactly at the resonance $|\boldsymbol{k}(q)|=1.6196\AA^{-1}$, and slightly less than the resonance and closer to the $K$-point $|\boldsymbol{k}(q)|=1.7005\AA^{-1}$. Each element of the quantum geometric tensor is found to exhibit a very distinct time dependence: The momentum components of the quantum metric $\left\{g_{xx},g_{xy},g_{yy}\right\}$ oscillate with time even after the pulse has passed, and increase with time in a nonlinear fashion; The temporal component $g_{tt}$ faithfully follows the time dependence of the pulse to display the same oscillation and envelope shape; The mixed component $g_{tx}$ and $g_{ty}$ also follow the pulse shape, but only at the resonant momentum do they increase significantly with time in a linear fashion after the pulse; The momentum component of the Berry curvature $\Omega_{xy}$ also oscillates and increases linearly with time; Finally, the mixed component Berry curvature $\Omega_{xt}$ and $\Omega_{yt}$ follow the pulse shape, and remain roughly constant after the pulse.

To understand the peculiar behavior of each component of the metric, in Appendix \ref{apx:driven_quantum_tunneling} we investigate a toy model of a chain of topological insulator in the presence of an oscillating mass term. The advantage of this toy model is that the time-dependent state $|\Psi(k,t)\rangle$ can be solved analytically under an approximation\cite{Grifoni98}, and so do the quantum metric $g_{\mu\nu}$ and Berry curvature $\Omega_{\mu\nu}$. We find that the time dependence of all the components shown in Fig.~\ref{fig:tqgt_gammaK} can be qualitatively captured by the toy model, indicating that the behavior shown in Fig.~\ref{fig:tqgt_gammaK} is fairly universal for time-dependent systems irrespective of the dimension and the drive. Finally, for completeness, in Appendix \ref{apx:quantum_geometry_whole_BZ} we show the dynamic quantum metric and Berry curvature in a wider range of momentum-time covering the whole Brillouin zone. The result shows that despite some weak responses throughout the Brillouin zone, the induced quantum metric and Berry curvature are the largest near the $K$ and $K'$ valleys.

\subsection{Fisher information waves \label{sec:fisher}}

The quantum metric of the static graphene has been proposed to be measurable by optical means\cite{vonGersdorff21_metric_curvature}, and so does the Berry curvature in other static solids\cite{Luu18,Chen22_dressed_Berry_metric,Kang25}. However, for the dynamic quantum metric and Berry curvature predicted in Sec.~\ref{sec:quantum_geometric_waves} that oscillates in the femtosecond scale, it is currently unclear to us how they can be detected experimentally. Nevertheless, in this section, we elaborate that the Fisher information contribution to the quantum metric can be readily detected by pump-probe experiments. The notion of Fisher information matrix arises when we treat the amplitudes in Eq.~(\ref{psikt_uc_uv}) as a probability mass function $\left\{|c_{1}|^{2},|c_{2}|^{2}\right\}=\left\{P_{1},P_{2}\right\}$ of finding the state in the two corresponding eigenstates. The Fisher information matrix that describes the variation of the probability mass function in the momentum-time manifold is defined as
\begin{equation}
    I_{\mu\nu}(\boldsymbol{k},t) = \sum_{i=1}^{2}P_i(\boldsymbol{k},t) \partial_{\mu}\ln P_i(\boldsymbol{k},t)\partial_{\nu}\ln P_i(\boldsymbol{k},t) = 4\sum_{i=1}^{2}\partial_{\mu}|c_i(\boldsymbol{k},t)|\,\partial_{\nu}|c_i(\boldsymbol{k},t)|
    \label{CFIM_definition}
\end{equation}
The dynamical quantum metric $g_{\mu\nu}$ is equivalent to a quarter of the Fisher information matrix $I_{\mu\nu}/4$ plus the corrections due to the phase of $c_i(\boldsymbol{k},t)$ and the momentum dependence of the basis states\cite{Facchi10}. Thus should the amplitudes $\left\{|c_{1}|^{2},|c_{2}|^2\right\}$ be measured experimentally, which can be quantified from the time-dependent spectral weight measured by tr-ARPES\cite{Gierz13}, the Fisher information matrix can be readily extracted through taking derivatives with respect to momentum and time according to Eq.~(\ref{CFIM_definition}).

\begin{figure}[ht]
    \centering
    \includegraphics[width=0.8\columnwidth]{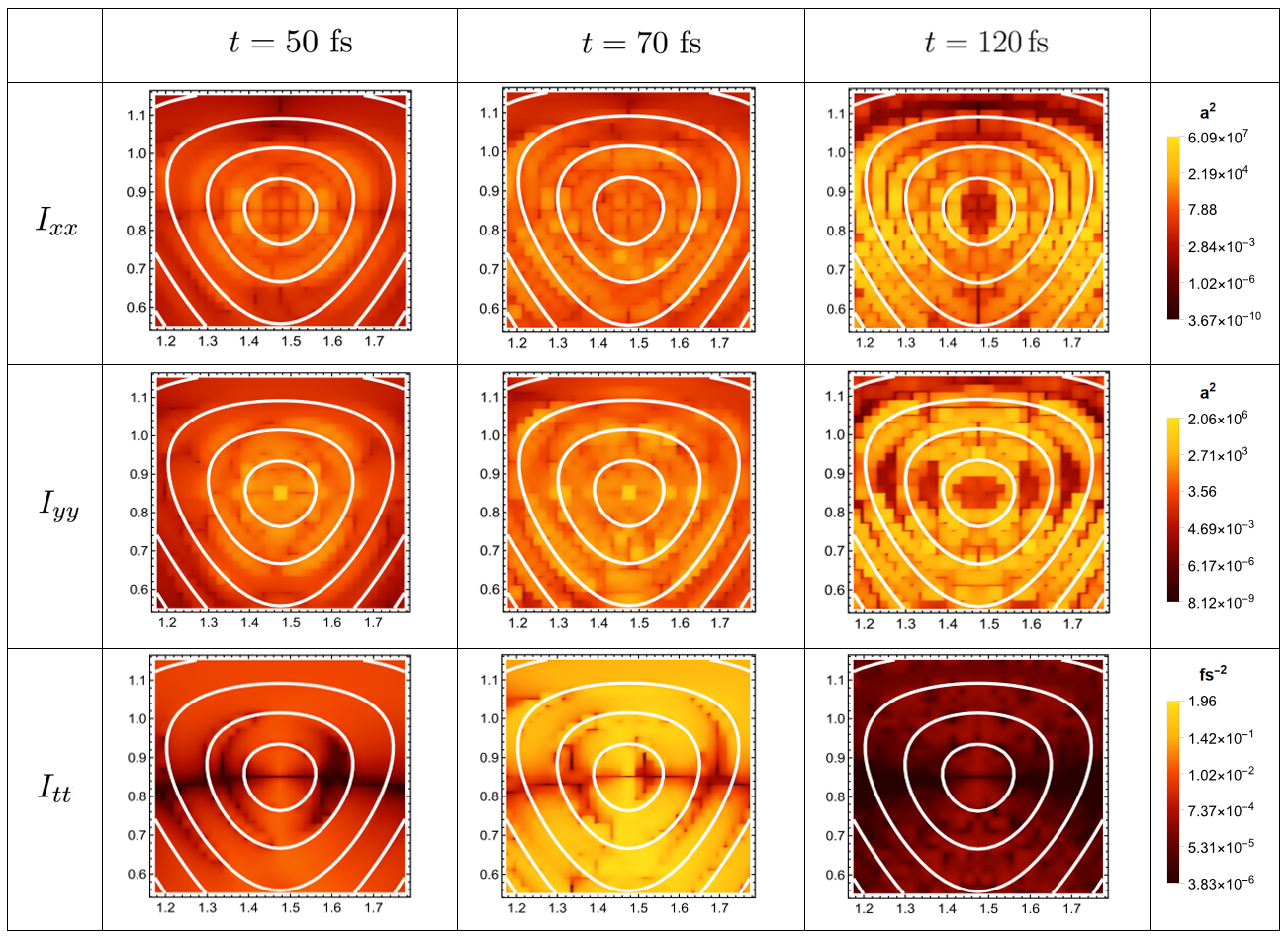}
    \caption{Momentum profile of the dynamic Fisher information matrix near the \(K\) valley caused by \(x\)-polarized pulse
evaluated at time \(t=50,70,120~\mathrm{fs}\).  We present the momentum components \(I_{xx}\) and
\(I_{yy}\) in units of \(a^{2}\), and the temporal component \(I_{tt}\) in units of
\(\mathrm{fs}^{-2}\). The white contours label the location of Floquet bands.}
    \label{fig:fisher_populations_rings}
\end{figure}

\begin{figure}[ht]
    \centering
    \includegraphics[width=0.8\columnwidth]{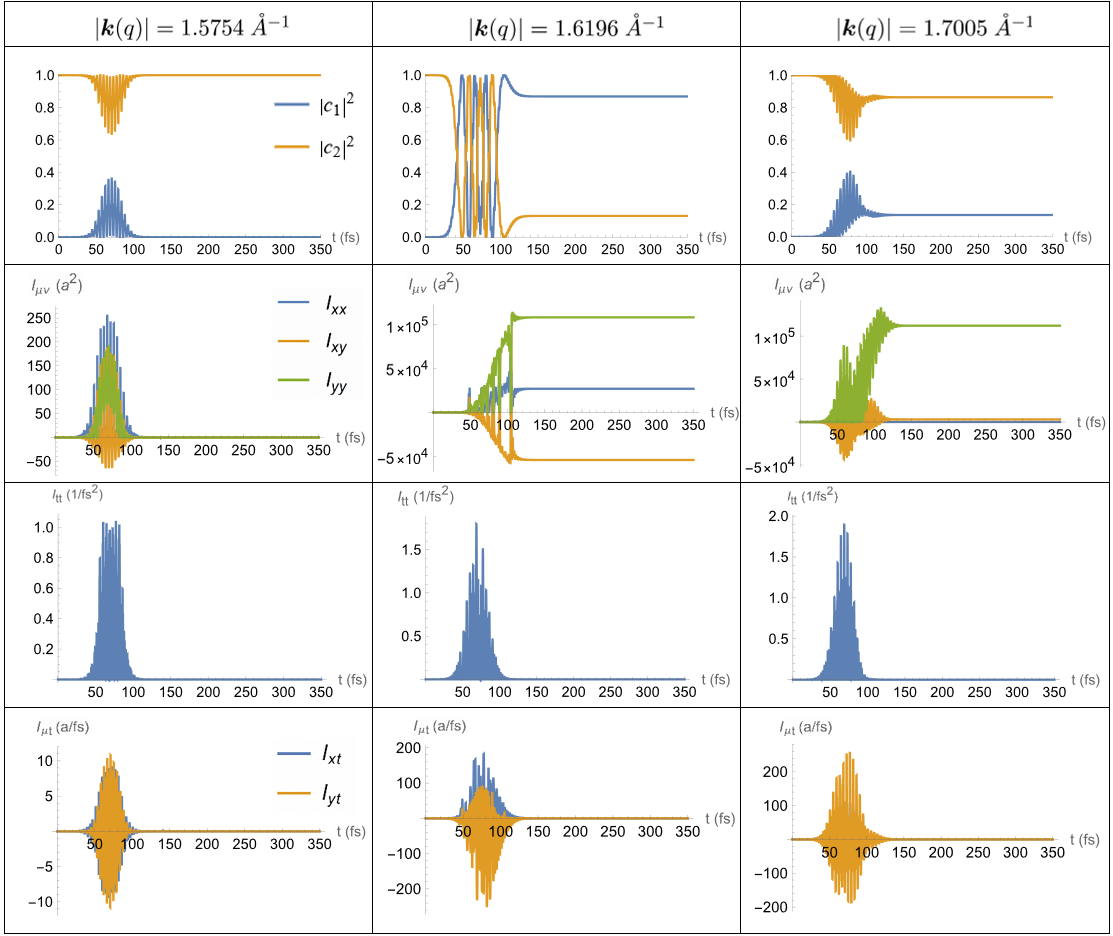}
    \caption{Time-varying band populations $\left\{|c_{1}|^{2},|c_{2}|^{2}\right\}$ (first row), momentum components $\left\{I_{xx},I_{xy},I_{yy}\right\}$ (second row), temporal component $I_{tt}$ (third row), and mixed component $\left\{I_{xt},I_{yt}\right\}$ of the dynamic Fisher information matrix along the \(\Gamma-K\) direction caused by \(x\)-polarized pulse. The columns correspond to momenta below, at, and above the
first resonant condition (see caption of Fig.~\ref{fig:tqgt_gammaK}). }
    \label{fig:fisher_populations}
\end{figure}

Numerical results for the elements of Fisher information matrix $\left\{I_{xx},I_{yy},I_{tt}\right\}$ near the $K$-point are shown in logarithmic scale in Fig.~\ref{fig:fisher_populations_rings}. The ring-like features that coincide with the Floquet bands are evident, of which we call Fisher information waves, suggesting that the Fisher information constructed from the amplitudes $\left\{|c_{1}|^{2},|c_{2}|^{2}\right\}$ also senses the formation of Floquet bands\cite{Sentef15}. The time dependence of the amplitudes and Fisher information matrix are shown in Fig.~\ref{fig:fisher_populations} for the same three momenta along the $\Gamma-K$ line as Fig.~\ref{fig:tqgt_gammaK}, manifesting the following features: The amplitudes $\left\{|c_{1}|^{2},|c_{2}|^{2}\right\}$ that characterize the population of the two bands oscillate rapidly during the pulse, signifying a fast transition between the two bands caused by the pulse; The momentum components $\left\{I_{xx},I_{yy},I_{xy}\right\}$ oscillate and increase during the pulse, until reaching a stable configuration after the pulse has passed; The temporal component $I_{tt}$ faithfully follows the pulse shape; The mixed components $\left\{I_{xt},I_{yt}\right\}$ are nonzero during the pulse, but do not completely follow the pulse shape. In Appendix \ref{apx:driven_quantum_tunneling}, we use the aforementioned toy model to analytically corroborate these peculiar behaviors, which suggest that they may be fairly universal in time-dependent systems regardless of the drive.

%Following the standard Fubini--Study construction of the quantum geometric tensor~\cite{Provost80} and its relation to the Fisher--Rao metric for pure states~\cite{Facchi10}, we use the following decomposition to interpret the driven two-band state. 

%Because the conduction-valence Bloch basis depends on momentum, the ordinary Fisher-plus-phase decomposition must be supplemented by the Berry connection of the static Bloch basis~\cite{WilczekZee1984}.

\section{Conclusions}
\label{sec:conclusions}

%{\cblue (1) Done here. }

In summary, by means of solving time-dependent Schr\"{o}dinger equation, we reveal that a short optical pulse causes the quantum metric of graphene in the 3D momentum-time manifold to become dynamic. The resulting quantum metric waves coincide with the Floquet bands caused by the pulse, and each component of the metric is found to have a very distinct time dependence that persists even after the pulse, which can be qualitatively explained by a toy model under the rotating wave approximation. Interestingly, the pulse also generates a Berry curvature wave that is otherwise absent in static graphene. The time dependence of the electron densities in the valence and conduction bands caused by the pulse, which can be viewed as the change of a probability mass function on the momentum-time manifold, further introduces a Fisher information wave that can be readily measured by pump-probe experiments.

We anticipate that the correlations and out-of-equilibrium effects that are not taken into account by our approach can significantly modify these dynamic quantum geometric properties in reality. In fact, when these complications are taken into account and Bloch states are ill-defined, it is even dubious how quantum metric and Berry curvature can be introduced. This profound question certainly deserves to be further investigated under realistic pump-probe experiment setup. Nevertheless, we anticipate that the time-dependent electron densities in the two bands, which have been extracted from the spectral function of tr-ARPES\cite{Gierz13}, would allow the Fisher information waves to be analyzed without ambiguity. Further comparison with experimental data is required to analyze whether our simple approach provides a reasonable description to the Fisher information waves in reality.

\section{Acknowledgments}

We acknowledge the support of the INCT project Advanced Quantum Materials, 
involving the Brazilian agencies CNPq (Proc. 408766/2024-7), FAPESP, and CAPES. W.C. acknowledges the financial support of the productivity in research fellowship from CNPq.

%\clearpage
%\appendix
%\renewcommand{\thesection}{S\arabic{section}}
%\renewcommand{\theHsection}{S\arabic{section}}
%\renewcommand{\thesubsection}{S\arabic{section}.\arabic{subsection}}
%\setcounter{section}{0}
%\setcounter{figure}{0}
%\setcounter{table}{0}
%\renewcommand{\thetable}{S\arabic{table}}
%\renewcommand{\thefigure}{S\arabic{figure}}
%\renewcommand{\theHfigure}{S\arabic{figure}}
%\setcounter{equation}{0}
%\renewcommand{\theequation}{S\arabic{equation}}
%\renewcommand{\theHequation}{S\arabic{equation}}

\appendix

\section{Driven quantum tunneling under rotating wave approximation \label{apx:driven_quantum_tunneling}}

Our aim in this section is to understand the time dependence of quantum metric, Berry curvature, and Fisher information matrix in Sec.~\ref{sec:quantum_geometric_waves} and \ref{sec:fisher} by means of a 1D toy model, whose quantum geometric properties can be solved analytically under rotating-wave approximation \cite{Grifoni98}. For this purpose, we consider the linearized Dirac Hamiltonian in 1D with an oscillating mass term ($\hbar=1$ thoughout this section)
\begin{eqnarray}
H(t)=M\sin\Omega t\;\sigma_{x}+k\sigma_{y}. 
\label{driven_1D_BDI_Mcost}
\end{eqnarray}
 Physically, this toy model starts as a topological semimetal phase $M=0$ at $t=0$, and the mass term $M$ oscillates between positive and negative values at $t>0$ with frequency $\Omega$. Although the periodically oscillating mass term is fundamentally different from the optical pulse described by Eq.~(\ref{Vt_pulse_shape}), we will see that it gives very similar quantum geometric properties. We expand the time-dependent state of interest by the eigenstates of the time-independent part $k\sigma_{y}$
\begin{eqnarray}
|\Psi(k,t)\rangle=c_{1}(t)e^{-ikt}\frac{1}{\sqrt{2}}\left(\begin{array}{c}
1 \\
i
\end{array}\right)+c_{2}(t)e^{ikt}\frac{1}{\sqrt{2}}\left(\begin{array}{c}
1 \\
-i
\end{array}\right),
\label{1D_class_BDI_psi_general}
\end{eqnarray}
with the amplitudes satisfying $|c_{1}(t)|^{2}+|c_{2}(t)|^{2}=1$. The time-dependent Schr\"{o}dinger equation gives 
\begin{eqnarray}
&&i\frac{d}{dt}\left(\begin{array}{c}
c_{1}(t)e^{-ikt}+c_{2}(t)e^{ikt} \\
ic_{1}(t)e^{-ikt}-ic_{2}(t)e^{ikt}
\end{array}\right)
\nonumber \\
&&=\left(\begin{array}{cc}
 & M\sin\Omega t-ik \\
M\sin\Omega t+ik & 
\end{array}\right)\left(\begin{array}{c}
c_{1}(t)e^{-ikt}+c_{2}(t)e^{ikt} \\
ic_{1}(t)e^{-ikt}-ic_{2}(t)e^{ikt}
\end{array}\right),
\label{1D_class_BDI_Schrodinger_matrix}
\end{eqnarray}
leading to the differential equations for the coefficients
\begin{eqnarray}
&&\frac{dc_{1}}{dt}=i\frac{M}{2}c_{2}\left\{e^{i(2k+\Omega)t}-e^{i(2k-\Omega)t}\right\},
\nonumber \\
&&\frac{dc_{2}}{dt}=i\frac{M}{2}c_{1}\left\{e^{-i(2k+\Omega)t}-e^{-i(2k-\Omega)t}\right\}.
\label{1D_class_BDI_dcdt_complex}
\end{eqnarray}
In general, these equations have to be solved numerically. Nevertheless, within the rotating wave approximation that drops the $e^{\pm i(2k+\Omega)t}$ terms based on the argument that these oscillations are too fast to induce a substantial transition between the two eigenstates, we can obtain some analytical results. Denoting $\delta=\Omega-2k$, the approximation amounts to 
\begin{eqnarray}
\frac{dc_{1}}{dt}=-i\frac{M}{2}c_{2}e^{-i\delta t},\;\;\;
\frac{dc_{2}}{dt}=-i\frac{M}{2}c_{1}e^{i\delta t}.
\label{1D_class_BDI_dcdt_rotating_wave}
\end{eqnarray}
We adopt the initial condition $c_{1}(0)=0$ and $c_{2}(0)=1$ such that all the particles are in the valence band at $t=0$. The solutions are
\begin{eqnarray}
&&c_{1}=-ie^{-i\delta t/2}\frac{M}{\Omega_{R}}\sin(\Omega_{R}t/2).
\nonumber \\
&&c_{2}=e^{i\delta t/2}\left[\cos(\Omega_{R}t/2)-\frac{i\delta}{\Omega_{R}}\sin(\Omega_{R}t/2)\right],
\label{1D_class_BDI_c1c2_solution}
\end{eqnarray}
where $\Omega_{R}=\sqrt{M^{2}+\delta^{2}}$ is the Rabi frequency.

The momentum-time quantum metric in this 2D manifold $k^{\mu}=(k,t)$ is defined from Eq.~(\ref{fidelity_expansion_metric}).
To calculate the quantum metric and Berry curvature, we write the state in Eq.~(\ref{1D_class_BDI_psi_general}) as $|\Psi\rangle = c_1(t) e^{-ikt} |u_{c}\rangle + c_2(t) e^{ikt} |u_{v}\rangle$. Both $|u_{c}\rangle$ and $|u_{v}\rangle$ are independent of $(k,t)$, so $\partial_\mu |\Psi\rangle
= \partial_\mu \big(c_1 e^{-ikt}\big)|u_{c}\rangle
+ \partial_\mu \big(c_2 e^{ikt}\big)|u_{v}\rangle$. Together with $\langle u_{\alpha}|u_{\beta}\rangle=\delta_{\alpha\beta}$, the quantum metric and Berry curvature can be calculated conveniently by
\begin{align}
&g_{\mu\nu}
= \frac{1}{2} \langle \partial_\mu \Psi | \partial_\nu \Psi \rangle
+ \frac{1}{2} \langle \partial_\nu \Psi | \partial_\mu \Psi \rangle
- \langle \partial_\mu \Psi | \Psi \rangle \langle \Psi | \partial_\nu \Psi \rangle
\\
&= \frac{1}{2} \partial_\mu(c_1^\ast e^{ikt}) \, \partial_\nu(c_1 e^{-ikt})
+ \frac{1}{2} \partial_\mu(c_2^\ast e^{-ikt}) \, \partial_\nu(c_2 e^{ikt})
\nonumber\\
&\quad + \frac{1}{2} \partial_\nu(c_1^\ast e^{ikt}) \, \partial_\mu(c_1 e^{-ikt})
+ \frac{1}{2} \partial_\nu(c_2^\ast e^{-ikt}) \, \partial_\mu(c_2 e^{ikt})
\nonumber\\
&- \Big[
\partial_\mu(c_1^\ast e^{ikt})(c_1 e^{-ikt})
+ \partial_\mu(c_2^\ast e^{-ikt})(c_2 e^{ikt})
\Big]
\Big[
(c_1^\ast e^{ikt}) \partial_\nu(c_1 e^{-ikt})
+ (c_2^\ast e^{-ikt}) \partial_\nu(c_2 e^{ikt})
\Big].
\nonumber \\
&\Omega_{\mu\nu}
= i \langle \partial_\mu \Psi | \partial_\nu \Psi \rangle
- i \langle \partial_\nu \Psi | \partial_\mu \Psi \rangle
\nonumber \\
&= i\, \partial_\mu(c_1^\ast e^{ikt}) \, \partial_\nu(c_1 e^{-ikt})
+ i\, \partial_\mu(c_2^\ast e^{-ikt}) \, \partial_\nu(c_2 e^{ikt})
\nonumber\\
& - i\, \partial_\nu(c_1^\ast e^{ikt}) \, \partial_\mu(c_1 e^{-ikt})
- i\, \partial_\nu(c_2^\ast e^{-ikt}) \, \partial_\mu(c_2 e^{ikt}).
\end{align}
By taking straightforward derivatives, these geometric quantities are found to be
\begin{align}
g_{kk}
&=
\frac{M^{2}}{2\Omega_{R}^{6}}
\Big(
3M^{2}
+2\delta^{2}\left(2+t^{2}\Omega_{R}^{2}\right)
-4\Omega_{R}^{2}\cos\left(\Omega_{R} t\right)
+M^{2}\cos\left(2\Omega_{R} t\right)
-4t\delta^{2}\Omega_{R}\sin\left(\Omega_{R} t\right)
\Big),
\nonumber \\
g_{tt}
&=
\frac{M^{2}}{8\Omega_{R}^{4}}
\Big(
2(4k^{2}+M^{2})^{2}
-8k(4k^{2}+M^{2})\Omega
+(16k^{2}+M^{2})\Omega^{2}
-8k\Omega^{3}
+2\Omega^{4}
\nonumber\\
&\qquad\qquad
+4\delta\Omega(4k^{2}+M^{2}-2k\Omega)
\cos\left(\Omega_{R} t\right)
-M^{2}\Omega^{2}
\cos\left(2\Omega_{R} t\right)
\Big),
\nonumber \\
g_{kt}
&=
\frac{M^{2}}{4\Omega_{R}^{6}}
\Big(
-2t\delta(4k^{2}+M^{2}-2k\Omega)\Omega_{R}^{2}
-2t\Omega\delta^{2}\Omega_{R}^{2}
\cos\left(\Omega_{R} t\right)
\nonumber\\
&\qquad\qquad
-2\Omega_{R}
\Big(
2k(4k^{2}+M^{2})
-2(6k^{2}+M^{2})\Omega
+6k\Omega^{2}
-\Omega^{3}
+M^{2}\Omega\cos\left(\Omega_{R} t\right)
\Big)
\sin\left(\Omega_{R} t\right)
\Big),
\nonumber \\
\Omega_{kt}
&=
\frac{M^{2}}{\Omega_{R}^{4}}
\Big(
-(4k^{2}+M^{2}-\Omega^{2})
\big(
-1+\cos(\Omega_{R} t)
\big)
-t\Omega\delta\Omega_{R}
\sin(\Omega_{R} t)
\Big).
\label{gtk_Omegatk_BDI_RWA}
\end{align}
From these expressions, we see that momentum component of the metric $g_{kk}$ oscillates and also grows quadratically with time, reminiscing the first row of Fig.~\ref{fig:tqgt_gammaK}; The temporal component $g_{tt}$ follows the oscillation of the drive, which coincides with the behavior in the second row of Fig.~\ref{fig:tqgt_gammaK}; The mixed component $g_{kt}$ oscillates and increases linearly with time, in agreement with the third row of Fig.~\ref{fig:tqgt_gammaK}; Finally, the mixed component Berry curvature $\Omega_{kt}$ oscillates and also has some component linear in time, which seems to capture some features in the fourth and fifth row of Fig.~\ref{fig:tqgt_gammaK}. Our investigation thus reveals a rather similar time dependence of quantum geometry in driven two-state systems, regardless of how it is driven.

%We proceed to treat $\left\{P_{1},P_{2}\right\}=\left\{|c_{1}|^{2},|c_{2}|^{2}\right\}$ as a probability mass function, since they represent the probability of finding the time-dependent quantum state in Eq.~(\ref{1D_class_BDI_psi_general}) to be in the basis states. We can then introduce the Fisher information matrix according to Eq.~(\ref{CFIM_definition}) to describe the information geometry of the momentum-time, leading to

The Fisher information matrix can be calculated in the same manner, yielding
\begin{align}
I_{kk}
&=
\frac{
4M^{2}\delta^{2}
\left(
4+t^{2}\Omega_{R}^{2}
+\left(-4+t^{2}\Omega_{R}^{2}\right)\cos(\Omega_{R} t)
-4t\Omega_{R}\sin(\Omega_{R} t)
\right)
}{
\Omega_{R}^{4}
\left(
M^{2}+2\delta^{2}+M^{2}\cos(\Omega_{R} t)
\right)
},
\nonumber \\
I_{tt}
&=
\frac{
2M^{2}\Omega_{R}^{2}
\cos^{2}\left(\Omega_{R} t/2\right)
}{
M^{2}+2\delta^{2}+M^{2}\cos(\Omega_{R} t)
},
\nonumber \\
I_{tk}
&=
-\frac{
2M^{2}\delta
\left(
\Omega_{R} t
+\Omega_{R} t \cos(\Omega_{R} t)
-2\sin(\Omega_{R} t)
\right)
}{
\Omega_{R}
\left(
M^{2}+2\delta^{2}+M^{2}\cos(\Omega_{R} t)
\right)
},
\label{Itk_BDI_RWA}
\end{align}
leading to a vanishing volume form $\sqrt{\det I_{\mu\nu}}=\sqrt{I_{tt}I_{kk}-I_{tk}^{2}}=0$. Our results indicate that the momentum component $I_{kk}$ oscillates and increases quadratically with time, which seems to coincide with the second row of Fig.~\ref{fig:fisher_populations} during the pulse period; The temporal component $I_{tt}$ simply follows the oscillation, which also reflects the behavior during the pulse in the third row of Fig.~\ref{fig:fisher_populations}; The mixed component $I_{kt}$ oscillates and linearly increases with time, reminiscing the peculiar shape in the fourth row of Fig.~\ref{fig:fisher_populations} that does not completely follow the pulse shape. Thus the information geometric properties of driven quantum systems may also display somewhat universal features irrespective of the details of the drive. 

%The quantum geometric properties in Eq.~(\ref{gtk_Omegatk_BDI_RWA}) and information geometric properties in Eq.~(\ref{Itk_BDI_RWA}) are highly nonlinear in momentum and time. While $\left\{g_{tt},I_{tt}\right\}$ has a rather regular oscillatory behavior, $\left\{g_{kk},g_{tk},I_{kk},I_{tk}\right\}$ grow with time and oscillate. Thus within the context of quantum geometry and information geometry, the momentum-time manifold is continuously vibrating and being stretched, despite the drive $M\sin\Omega t$ is steady and periodic. {\cblue Comment on whether they are consistent with the behavior in graphene}.

%\section*{Supplemental Material}
%\addcontentsline{toc}{section}{Supplemental Material}

\begin{figure}[h]
    \centering
    \includegraphics[width=0.8\columnwidth]{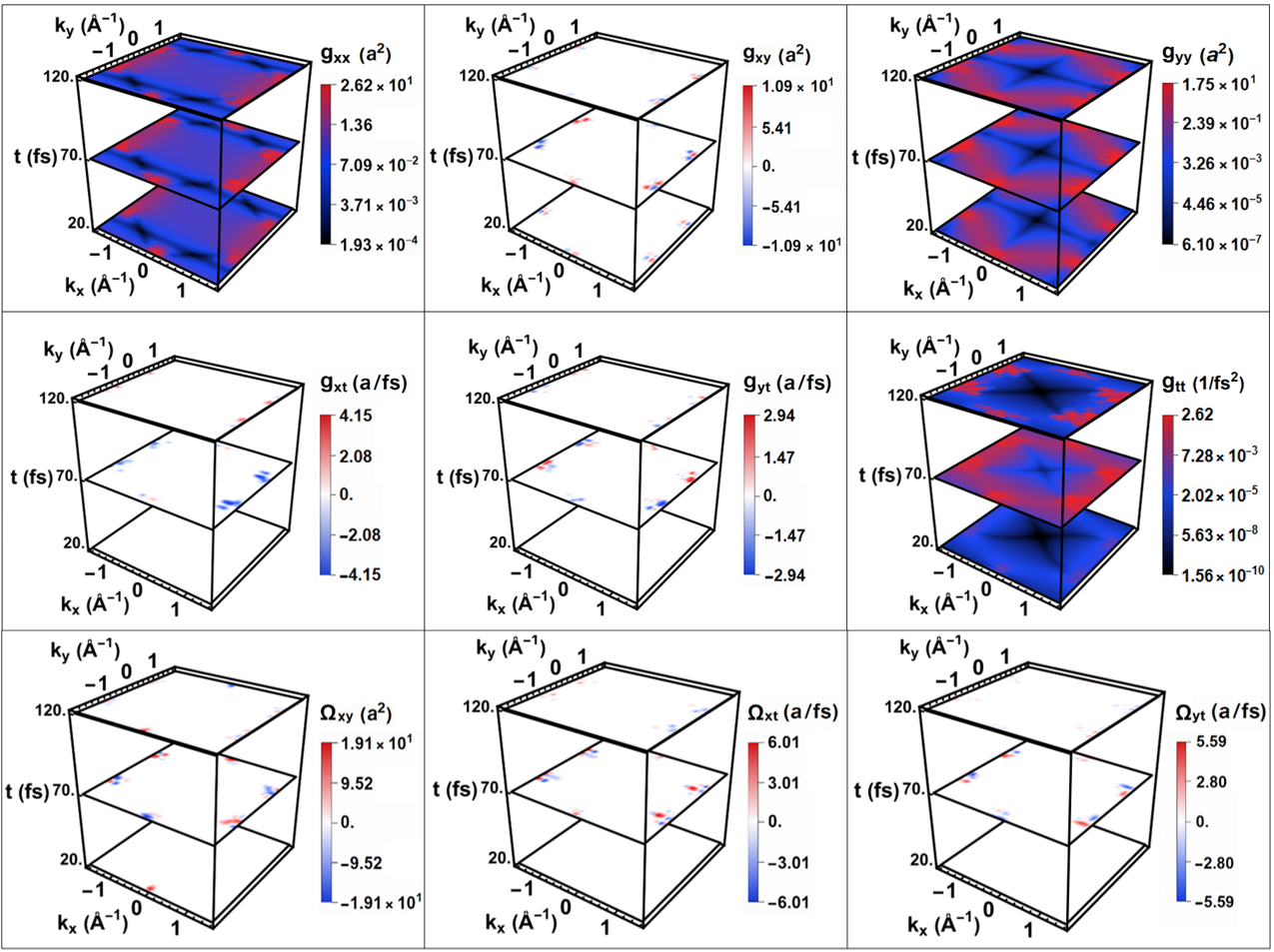}
    \caption{The momentum-time pattern of the components of the dynamic quantum geometric tensor in a wider momentum space that covers the first Brillouin zone, evaluated at three time slices representing before the pulse maximum $t=20$fs, peak of the pulse $t=70$fs, and after the pulse $t=120$fs. The components are labeled at each panel. }
    \label{fig:S_global_metric}
\end{figure}

\section{Dynamic quantum geometry in the whole Brillouin zone \label{apx:quantum_geometry_whole_BZ}}

In Figure~\ref{fig:S_global_metric}, we present all the components of the quantum metric and Berry curvature in the whole Brillouin zone, at three time slices corresponding to the beginning of the pulse $t=20$fs, peak of the pulse $70$fs, and right after the pulse $120$fs. While there is some contribution throughout the whole Brillouin zone, the figure indicates that the maxima of the quantum metric and Berry curvature are always located around the $K$ and $K'$ valleys.

\bibliographystyle{unsrt}
\bibliography{bibliography}

@article{Berry84,
	author = {Berry, M. V.},
	doi = {10.1098/rspa.1984.0023},
	issn = {0080-4630},
	journal = {Proc. R. Soc. A},
	number = {1802},
	pages = {45--57},
	publisher = {The Royal Society},
	title = {Quantal Phase Factors Accompanying Adiabatic Changes},
	url = {http://rspa.royalsocietypublishing.org/content/392/1802/45},
	volume = {392},
	year = {1984}}

@article{Xiao10,
	author = {Xiao, Di and Chang, Ming-Che and Niu, Qian},
	doi = {10.1103/RevModPhys.82.1959},
	issue = {3},
	journal = {Rev. Mod. Phys.},
	month = jul,
	numpages = {0},
	pages = {1959--2007},
	publisher = {American Physical Society},
	title = {Berry phase effects on electronic properties},
	volume = {82},
	year = {2010}}

@article{vonGersdorff21_metric_curvature,
	author = {von Gersdorff, Gero and Chen, Wei},
	doi = {10.1103/PhysRevB.104.195133},
	issue = {19},
	journal = {Phys. Rev. B},
	month = {Nov},
	numpages = {7},
	pages = {195133},
	publisher = {American Physical Society},
	title = {Measurement of topological order based on metric-curvature correspondence},
	url = {https://link.aps.org/doi/10.1103/PhysRevB.104.195133},
	volume = {104},
	year = {2021}}

@Article{Chen22_dressed_Berry_metric,
	title={Measurement of interaction-dressed Berry curvature and quantum metric in  solids by optical absorption},
	author={Wei Chen and Gero von Gersdorff},
	journal={SciPost Phys. Core},
	volume={5},
	pages={040},
	year={2022},
	publisher={SciPost},
	doi={10.21468/SciPostPhysCore.5.3.040},
	url={https://scipost.org/10.21468/SciPostPhysCore.5.3.040},
}

@Article{Gierz13,
author={Gierz, Isabella
and Petersen, Jesse C.
and Mitrano, Matteo
and Cacho, Cephise
and Turcu, I. C. Edmond
and Springate, Emma
and St\"{o}hr, Alexander
and K\"{o}hler, Axel
and Starke, Ulrich
and Cavalleri, Andrea},
title={Snapshots of non-equilibrium Dirac carrier distributions in graphene},
journal={Nat. Mater.},
year={2013},
month={Dec},
day={01},
volume={12},
number={12},
pages={1119-1124},
issn={1476-4660},
doi={10.1038/nmat3757},
url={https://doi.org/10.1038/nmat3757}
}

@article{Provost80,
author = "Provost, J. P. and Vallee, G.",
fjournal = "Communications in Mathematical Physics",
journal = "Comm. Math. Phys.",
number = "3",
pages = "289--301",
publisher = "Springer",
title = "Riemannian structure on manifolds of quantum states",
url = "https://projecteuclid.org:443/euclid.cmp/1103908308",
volume = "76",
year = "1980"
}

@article{Ozawa18,
  title = {Extracting the quantum metric tensor through periodic driving},
  author = {Ozawa, Tomoki and Goldman, Nathan},
  journal = {Phys. Rev. B},
  volume = {97},
  issue = {20},
  pages = {201117},
  numpages = {6},
  year = {2018},
  month = {May},
  publisher = {American Physical Society},
  doi = {10.1103/PhysRevB.97.201117},
  url = {https://link.aps.org/doi/10.1103/PhysRevB.97.201117}
}

@Article{Ahn22,
author={Ahn, Junyeong
and Guo, Guang-Yu
and Nagaosa, Naoto
and Vishwanath, Ashvin},
title={Riemannian geometry of resonant optical responses},
journal={Nat. Phys.},
year={2022},
month={Mar},
day={01},
volume={18},
number={3},
pages={290-295},
issn={1745-2481},
doi={10.1038/s41567-021-01465-z},
url={https://doi.org/10.1038/s41567-021-01465-z}
}

@Article{Komissarov24,
author={Komissarov, Ilia
and Holder, Tobias
and Queiroz, Raquel},
title={The quantum geometric origin of capacitance in insulators},
journal={Nature Communications},
year={2024},
month={May},
day={30},
volume={15},
number={1},
pages={4621},
issn={2041-1723},
doi={10.1038/s41467-024-48808-x},
url={https://doi.org/10.1038/s41467-024-48808-x}
}

@article{Chen25_optical_marker,
  title = {Dielectric and optical markers originating from quantum geometry},
  author = {Chen, Wei},
  journal = {Phys. Rev. B},
  volume = {111},
  issue = {8},
  pages = {085202},
  numpages = {12},
  year = {2025},
  month = {Feb},
  publisher = {American Physical Society},
  doi = {10.1103/PhysRevB.111.085202},
  url = {https://link.aps.org/doi/10.1103/PhysRevB.111.085202}
}

@article{Grandi10,
  title = {Quench dynamics near a quantum critical point},
  author = {De Grandi, C. and Gritsev, V. and Polkovnikov, A.},
  journal = {Phys. Rev. B},
  volume = {81},
  issue = {1},
  pages = {012303},
  numpages = {4},
  year = {2010},
  month = {Jan},
  publisher = {American Physical Society},
  doi = {10.1103/PhysRevB.81.012303},
  url = {https://link.aps.org/doi/10.1103/PhysRevB.81.012303}
}

@article{Grandi10_2,
  title = {Quench dynamics near a quantum critical point: Application to the sine-Gordon model},
  author = {De Grandi, C. and Gritsev, V. and Polkovnikov, A.},
  journal = {Phys. Rev. B},
  volume = {81},
  issue = {22},
  pages = {224301},
  numpages = {21},
  year = {2010},
  month = {Jun},
  publisher = {American Physical Society},
  doi = {10.1103/PhysRevB.81.224301},
  url = {https://link.aps.org/doi/10.1103/PhysRevB.81.224301}
}

@article{Jafari20,
  title = {Dynamics of quantum coherence and quantum Fisher information after a sudden quench},
  author = {Jafari, R. and Akbari, Alireza},
  journal = {Phys. Rev. A},
  volume = {101},
  issue = {6},
  pages = {062105},
  numpages = {9},
  year = {2020},
  month = {Jun},
  publisher = {American Physical Society},
  doi = {10.1103/PhysRevA.101.062105},
  url = {https://link.aps.org/doi/10.1103/PhysRevA.101.062105}
}

@misc{Wu24,
      title={Universal critical dynamics of quantum geometry}, 
      author={Shuohang Wu and Zhengxin Guo and Zijian Xiong and Yuan Yao and Zi Cai},
      year={2024},
      eprint={2401.17885},
      archivePrefix={arXiv},
      primaryClass={cond-mat.quant-gas},
      url={https://arxiv.org/abs/2401.17885},
      note= {arXiv:2401.17885},
}

@misc{Diaz25,
      title={Time-dependent quantum geometric tensor and some applications}, 
      author={Bogar Díaz and Diego Gonzalez and Marcos J. Hernández and J. David Vergara},
      year={2025},
      eprint={2502.01788},
      archivePrefix={arXiv},
      primaryClass={quant-ph},
      url={https://arxiv.org/abs/2502.01788},
      note= {arXiv:2502.01788},
}

@Article{Pang17,
author={Pang, Shengshi
and Jordan, Andrew N.},
title={Optimal adaptive control for quantum metrology with time-dependent Hamiltonians},
journal={Nature Communications},
year={2017},
month={Mar},
day={09},
volume={8},
number={1},
pages={14695},
issn={2041-1723},
doi={10.1038/ncomms14695},
url={https://doi.org/10.1038/ncomms14695}
}

@article{Grifoni98,
title = {Driven quantum tunneling},
journal = {Physics Reports},
volume = {304},
number = {5},
pages = {229-354},
year = {1998},
issn = {0370-1573},
doi = {https://doi.org/10.1016/S0370-1573(98)00022-2},
url = {https://www.sciencedirect.com/science/article/pii/S0370157398000222},
author = {Milena Grifoni and Peter Hänggi}
}

@article{Gierz15,
  title = {Tracking Primary Thermalization Events in Graphene with Photoemission at Extreme Time Scales},
  author = {Gierz, I. and Calegari, F. and Aeschlimann, S. and Ch\'avez Cervantes, M. and Cacho, C. and Chapman, R. T. and Springate, E. and Link, S. and Starke, U. and Ast, C. R. and Cavalleri, A.},
  journal = {Phys. Rev. Lett.},
  volume = {115},
  issue = {8},
  pages = {086803},
  numpages = {5},
  year = {2015},
  month = {Aug},
  publisher = {American Physical Society},
  doi = {10.1103/PhysRevLett.115.086803},
  url = {https://link.aps.org/doi/10.1103/PhysRevLett.115.086803}
}

@article{Johannsen13,
  title = {Direct View of Hot Carrier Dynamics in Graphene},
  author = {Johannsen, Jens Christian and Ulstrup, S\o{}ren and Cilento, Federico and Crepaldi, Alberto and Zacchigna, Michele and Cacho, Cephise and Turcu, I. C. Edmond and Springate, Emma and Fromm, Felix and Raidel, Christian and Seyller, Thomas and Parmigiani, Fulvio and Grioni, Marco and Hofmann, Philip},
  journal = {Phys. Rev. Lett.},
  volume = {111},
  issue = {2},
  pages = {027403},
  numpages = {5},
  year = {2013},
  month = {Jul},
  publisher = {American Physical Society},
  doi = {10.1103/PhysRevLett.111.027403},
  url = {https://link.aps.org/doi/10.1103/PhysRevLett.111.027403}
}

@Article{Wagner14,
author={Wagner, Martin
and Fei, Zhe
and McLeod, Alexander S.
and Rodin, Aleksandr S.
and Bao, Wenzhong
and Iwinski, Eric G.
and Zhao, Zeng
and Goldflam, Michael
and Liu, Mengkun
and Dominguez, Gerardo
and Thiemens, Mark
and Fogler, Michael M.
and Castro Neto, Antonio H.
and Lau, Chun Ning
and Amarie, Sergiu
and Keilmann, Fritz
and Basov, D. N.},
title={Ultrafast and Nanoscale Plasmonic Phenomena in Exfoliated Graphene Revealed by Infrared Pump--Probe Nanoscopy},
journal={Nano Lett.},
year={2014},
month={Feb},
day={12},
publisher={American Chemical Society},
volume={14},
number={2},
pages={894-900},
issn={1530-6984},
doi={10.1021/nl4042577},
url={https://doi.org/10.1021/nl4042577}
}

@Article{Sentef15,
author={Sentef, M. A.
and Claassen, M.
and Kemper, A. F.
and Moritz, B.
and Oka, T.
and Freericks, J. K.
and Devereaux, T. P.},
title={Theory of Floquet band formation and local pseudospin textures in pump-probe photoemission of graphene},
journal={Nature Communications},
year={2015},
month={May},
day={11},
volume={6},
number={1},
pages={7047},
issn={2041-1723},
doi={10.1038/ncomms8047},
url={https://doi.org/10.1038/ncomms8047}
}

@Article{Strait11,
author={Strait, Jared H.
and Wang, Haining
and Shivaraman, Shriram
and Shields, Virgil
and Spencer, Michael
and Rana, Farhan},
title={Very Slow Cooling Dynamics of Photoexcited Carriers in Graphene Observed by Optical-Pump Terahertz-Probe Spectroscopy},
journal={Nano Lett.},
year={2011},
month={Nov},
day={09},
publisher={American Chemical Society},
volume={11},
number={11},
pages={4902-4906},
issn={1530-6984},
doi={10.1021/nl202800h},
url={https://doi.org/10.1021/nl202800h}
}

@article{Breusing11,
  title = {Ultrafast nonequilibrium carrier dynamics in a single graphene layer},
  author = {Breusing, M. and Kuehn, S. and Winzer, T. and Mali\'{c}, E. and Milde, F. and Severin, N. and Rabe, J. P. and Ropers, C. and Knorr, A. and Elsaesser, T.},
  journal = {Phys. Rev. B},
  volume = {83},
  issue = {15},
  pages = {153410},
  numpages = {4},
  year = {2011},
  month = {Apr},
  publisher = {American Physical Society},
  doi = {10.1103/PhysRevB.83.153410},
  url = {https://link.aps.org/doi/10.1103/PhysRevB.83.153410}
}

@article{Dawlaty08,
    author = {Dawlaty, Jahan M. and Shivaraman, Shriram and Chandrashekhar, Mvs and Rana, Farhan and Spencer, Michael G.},
    title = {Measurement of ultrafast carrier dynamics in epitaxial graphene},
    journal = {Appl. Phys. Lett.},
    volume = {92},
    number = {4},
    pages = {042116},
    year = {2008},
    month = {01},
    issn = {0003-6951},
    doi = {10.1063/1.2837539},
    url = {https://doi.org/10.1063/1.2837539},
    eprint = {https://pubs.aip.org/aip/apl/article-pdf/doi/10.1063/1.2837539/14386809/042116_1_online.pdf},
}

@Article{Plotzing14,
author={Pl{\"o}tzing, T.
and Winzer, T.
and Malic, E.
and Neumaier, D.
and Knorr, A.
and Kurz, H.},
title={Experimental Verification of Carrier Multiplication in Graphene},
journal={Nano Lett.},
year={2014},
month={Sep},
day={10},
publisher={American Chemical Society},
volume={14},
number={9},
pages={5371-5375},
issn={1530-6984},
doi={10.1021/nl502114w},
url={https://doi.org/10.1021/nl502114w}
}

@Article{Brida13,
author={Brida, D.
and Tomadin, A.
and Manzoni, C.
and Kim, Y. J.
and Lombardo, A.
and Milana, S.
and Nair, R. R.
and Novoselov, K. S.
and Ferrari, A. C.
and Cerullo, G.
and Polini, M.},
title={Ultrafast collinear scattering and carrier multiplication in graphene},
journal={Nat. Commun.},
year={2013},
month={Jun},
day={17},
volume={4},
number={1},
pages={1987},
issn={2041-1723},
doi={10.1038/ncomms2987},
url={https://doi.org/10.1038/ncomms2987}
}

@article{Wang10,
    author = {Wang, Haining and Strait, Jared H. and George, Paul A. and Shivaraman, Shriram and Shields, Virgil B. and Chandrashekhar, Mvs and Hwang, Jeonghyun and Rana, Farhan and Spencer, Michael G. and Ruiz-Vargas, Carlos S. and Park, Jiwoong},
    title = {Ultrafast relaxation dynamics of hot optical phonons in graphene},
    journal = {Appl. Phys. Lett.},
    volume = {96},
    number = {8},
    pages = {081917},
    year = {2010},
    month = {02},
    issn = {0003-6951},
    doi = {10.1063/1.3291615},
    url = {https://doi.org/10.1063/1.3291615},
    eprint = {https://pubs.aip.org/aip/apl/article-pdf/doi/10.1063/1.3291615/14427573/081917_1_online.pdf},
}

@article{Winnerl13,
doi = {10.1088/0953-8984/25/5/054202},
url = {https://doi.org/10.1088/0953-8984/25/5/054202},
year = {2013},
month = {jan},
publisher = {IOP Publishing},
volume = {25},
number = {5},
pages = {054202},
author = {Winnerl, S and Göttfert, F and Mittendorff, M and Schneider, H and Helm, M and Winzer, T and Malic, E and Knorr, A and Orlita, M and Potemski, M and Sprinkle, M and Berger, C and de Heer, W A},
title = {Time-resolved spectroscopy on epitaxial graphene in the infrared spectral range: relaxation dynamics and saturation behavior},
journal = {J. Phys. Condens. Matter},
}

@article{Winnerl11,
  title = {Carrier Relaxation in Epitaxial Graphene Photoexcited Near the Dirac Point},
  author = {Winnerl, S. and Orlita, M. and Plochocka, P. and Kossacki, P. and Potemski, M. and Winzer, T. and Malic, E. and Knorr, A. and Sprinkle, M. and Berger, C. and de Heer, W. A. and Schneider, H. and Helm, M.},
  journal = {Phys. Rev. Lett.},
  volume = {107},
  issue = {23},
  pages = {237401},
  numpages = {5},
  year = {2011},
  month = {Nov},
  publisher = {American Physical Society},
  doi = {10.1103/PhysRevLett.107.237401},
  url = {https://link.aps.org/doi/10.1103/PhysRevLett.107.237401}
}

@Article{Tielrooij13,
author={Tielrooij, K. J.
and Song, J. C. W.
and Jensen, S. A.
and Centeno, A.
and Pesquera, A.
and Zurutuza Elorza, A.
and Bonn, M.
and Levitov, L. S.
and Koppens, F. H. L.},
title={Photoexcitation cascade and multiple hot-carrier generation in graphene},
journal={Nat. Phys.},
year={2013},
month={Apr},
day={01},
volume={9},
number={4},
pages={248-252},
issn={1745-2481},
doi={10.1038/nphys2564},
url={https://doi.org/10.1038/nphys2564}
}

@Article{Merboldt25,
author={Merboldt, Marco
and Sch{\"u}ler, Michael
and Schmitt, David
and Bange, Jan Philipp
and Bennecke, Wiebke
and Gadge, Karun
and Pierz, Klaus
and Schumacher, Hans Werner
and Momeni, Davood
and Steil, Daniel
and Manmana, Salvatore R.
and Sentef, Michael A.
and Reutzel, Marcel
and Mathias, Stefan},
title={Observation of Floquet states in graphene},
journal={Nat. Phys.},
year={2025},
month={Jul},
day={01},
volume={21},
number={7},
pages={1093-1099},
issn={1745-2481},
doi={10.1038/s41567-025-02889-7},
url={https://doi.org/10.1038/s41567-025-02889-7}
}

@article{Schuler21,
title = {Theory of subcycle time-resolved photoemission: Application to terahertz photodressing in graphene},
journal = {J. Electron Spectrosc. Relat. Phenom.},
volume = {253},
pages = {147121},
year = {2021},
issn = {0368-2048},
doi = {https://doi.org/10.1016/j.elspec.2021.147121},
url = {https://www.sciencedirect.com/science/article/pii/S0368204821000736},
author = {Michael Sch\"{u}ler and Michael A. Sentef},
}

@article{Fisher25, 
title={Theory of Statistical Estimation}, 
volume={22}, 
DOI={10.1017/S0305004100009580}, 
number={5}, 
journal={Math. Proc. Camb. Philos. Soc.}, 
author={Fisher, R. A.}, 
year={1925}, 
pages={700–725}}

@article{Facchi10,
title = {Classical and quantum Fisher information in the geometrical formulation of quantum mechanics},
journal = {Physics Letters A},
volume = {374},
number = {48},
pages = {4801-4803},
year = {2010},
issn = {0375-9601},
doi = {https://doi.org/10.1016/j.physleta.2010.10.005},
url = {https://www.sciencedirect.com/science/article/pii/S0375960110013204},
author = {Paolo Facchi and Ravi Kulkarni and V.I. Man'ko and Giuseppe Marmo and E.C.G. Sudarshan and Franco Ventriglia},
}

@article{Kochan17,
  title = {Model spin-orbit coupling Hamiltonians for graphene systems},
  author = {Kochan, Denis and Irmer, Susanne and Fabian, Jaroslav},
  journal = {Phys. Rev. B},
  volume = {95},
  issue = {16},
  pages = {165415},
  numpages = {19},
  year = {2017},
  month = {Apr},
  publisher = {American Physical Society},
  doi = {10.1103/PhysRevB.95.165415},
  url = {https://link.aps.org/doi/10.1103/PhysRevB.95.165415}
}

@Book{Amari16,
  title      = {Information Geometry and Its Applications},
  publisher  = {Springer},
  year       = {2016},
  author     = {Amari, S.},
  month      = feb,
  isbn       = { ‎978-4431559771},
  totalpages = {387},
}

@article{CastroNeto09,
  title = {The electronic properties of graphene},
  author = {Castro Neto, A. H. and Guinea, F. and Peres, N. M. R. and Novoselov, K. S. and Geim, A. K.},
  journal = {Rev. Mod. Phys.},
  volume = {81},
  issue = {1},
  pages = {109--162},
  numpages = {0},
  year = {2009},
  month = {Jan},
  publisher = {American Physical Society},
  doi = {10.1103/RevModPhys.81.109},
  url = {https://link.aps.org/doi/10.1103/RevModPhys.81.109}
}

@Article{Luu18,
author={Luu, Tran Trung
and W{\"o}rner, Hans Jakob},
title={Measurement of the Berry curvature of solids using high-harmonic spectroscopy},
journal={Nat. Commun.},
year={2018},
month={Mar},
day={02},
volume={9},
number={1},
pages={916},
issn={2041-1723},
doi={10.1038/s41467-018-03397-4},
url={https://doi.org/10.1038/s41467-018-03397-4}
}

@Article{Kang25,
author={Kang, Mingu
and Kim, Sunje
and Qian, Yuting
and Neves, Paul M.
and Ye, Linda
and Jung, Junseo
and Puntel, Denny
and Mazzola, Federico
and Fang, Shiang
and Jozwiak, Chris
and Bostwick, Aaron
and Rotenberg, Eli
and Fuji, Jun
and Vobornik, Ivana
and Park, Jae-Hoon
and Checkelsky, Joseph G.
and Yang, Bohm-Jung
and Comin, Riccardo},
title={Measurements of the quantum geometric tensor in solids},
journal={Nat. Phys.},
year={2025},
month={Jan},
day={01},
volume={21},
number={1},
pages={110-117},
issn={1745-2481},
doi={10.1038/s41567-024-02678-8},
url={https://doi.org/10.1038/s41567-024-02678-8}
}

\end{document}